\documentclass[showaps,graphicx,twocolumn]{revtex4}
\usepackage{amssymb}
\usepackage{amsmath}
\usepackage{graphicx}
\usepackage{array}
\usepackage{multirow}
\usepackage[colorlinks,
linkcolor=blue,
anchorcolor=blue,
citecolor=blue,
urlcolor=blue
]{hyperref}

\begin{document}

\title{Linear-optical four-dimensional Bell state measurement with two-photon interference}

\author{Zhi Zeng$^{1,2,}$\footnote{Corresponding author:
		zengzhiphy@yeah.net}}
\affiliation{{$^1$CAS Key Laboratory of Quantum Optics and Center of Cold Atom Physics, Shanghai Institute of Optics and Fine Mechanics, Chinese Academy of Sciences, Shanghai 201800, China\\}
{$^2$Wangzhijiang Innovation Center for Laser, Aerospace Laser Technology and System Department, Shanghai Institute of Optics and Fine Mechanics, Chinese Academy of Sciences, Shanghai 201800, China}}

\begin{abstract}
We theoretically investigate the distinguishability of a set of mutually orthogonal four-dimensional Bell states of photon system in path degree of freedom using only linear optics, resorting to the two-photon interference. With quantum interference effect, we find that the 16 four-dimensional Bell states can be classified into 7 groups, which can support the transmission of $log_27=2.81$ bits classical information with just sending one photon in the quantum superdense coding protocol. When an auxiliary two-dimensional polarization entanglement is introduced, the 16 four-dimensional Bell states then can be classified into 12 groups, which can promote the channel capacity to $log_212=3.58$ bits with encoding one photon. Our results are significant for photonic superdense coding, and can be useful for other quantum information technologies involved linear-optical high-dimensional Bell state measurement. 
\end{abstract}

\maketitle

\section{Introduction}
High-dimensional quantum entanglement has been widely researched in the past years \cite{1}, and has also played an important role in the high-capacity quantum information technology (QIT) \cite{2}. The Bell state measurement (BSM) of two-qubit system is an indispensable procedure for many QITs, such as quantum dense coding \cite{3}, quantum teleportation \cite{4}, and quantum entanglement swapping \cite{5}. However, the optimal strategy for linear-optical BSM of two-dimensional quantum system is to divide the four orthogonal Bell states into three groups, and only two states can be determined, which leads to a 50\% success probability \cite{6,7}. For the high-dimensional BSM (HDBSM) of two-qudit system, it has been proven that the complete discrimination of a set of orthogonal high-dimensional Bell basis is also impossible if only linear optics is utilized \cite{8}. In recent quantum optical experiments, the linear-optical HDBSM has been exploited for high-dimensional quantum communication, and the not high distinguishing efficiency of HDBSM can directly decrease the communication efficiency \cite{9,10,11,12}. Therefore, the research of linear-optical HDBSM not only is a fundamental problem in quantum information theory, but also can be valuable for practical applications. 

In this paper, we investigate the linear-optical distinguishability of photonic four-dimensional Bell state in path degree of freedom (DOF) based on the two-photon interference, which are now popularly termed as the Hong-Ou-Mandel (HOM) effect \cite{13,14,15,16}. As a particular phenomenon in quantum world with absolutely no analogue in classical physics, two-photon interference has already been shown to be helpful for traditional two-dimensional BSM and quantum dense coding \cite{17,18,19,20,21,22}. However, compared with two-dimensional quantum system, the high-dimensional quantum system becomes more complicated. By using the two-photon interference and simple linear optical element, we find that the 16 four-dimensional path Bell states can be separated into 7 groups. When an auxiliary two-dimensional Bell state in polarization DOF is introduced, the 16 states then can be separated into 12 groups. These two linear-optical HDBSM approaches can be utilized for superdense coding with four-dimensional Bell state, and be able to support the channel capacities of $log_27=2.81$ bits and $log_212=3.58$ bits, respectively. Our approaches for HDBSM are important for quantum measurement of entangled state in high dimensions, and will be useful for other HDBSM-based QITs with quantum interference.

\section{Linear-optical HDBSM for superdense coding using two-photon interference}
The form of 16 mutually orthogonal four-dimensional path Bell states of two-photon system can be written as
\begin{eqnarray}
|\psi_{0n}^{m}\rangle = \frac{1}{2}(|00\rangle + e^{in\pi}|11\rangle +(-1)^m |22\rangle +(-1)^m e^{in\pi}|33\rangle)_{AB}, \nonumber  \\
|\psi_{1n}^{m}\rangle = \frac{1}{2}(|01\rangle + e^{in\pi}|10\rangle +(-1)^m |23\rangle +(-1)^m e^{in\pi}|32\rangle)_{AB}, \nonumber  \\
|\psi_{2n}^{m}\rangle = \frac{1}{2}(|02\rangle + e^{in\pi}|13\rangle +(-1)^m |20\rangle +(-1)^m e^{in\pi}|31\rangle)_{AB}, \nonumber  \\
|\psi_{3n}^{m}\rangle = \frac{1}{2}(|03\rangle + e^{in\pi}|12\rangle +(-1)^m |21\rangle +(-1)^m e^{in\pi}|30\rangle)_{AB}.  \nonumber  \\
\end{eqnarray}
Here $m,n=0,1$, and $|0\rangle$, $|1\rangle$, $|2\rangle$, and $|3\rangle$ are four different quantum states of the entangled photons $AB$ in path DOF. In superdense coding, the receiver first prepares the four-dimensional path Bell state $|\psi_{00}^{0}\rangle$, and sends one of the photons to the sender with quantum channel. After receiving photon, the sender performs one of the 16 unitary operations $U^{m}_{jn}(j=0,1,2,3)$ on his photon, which can realize the transformation between the 16 four-dimensional path Bell states to encode classical information. The operations can be expressed as
\begin{eqnarray}
U_{0n}^{m} = |0\rangle\langle0| + e^{in\pi}|1\rangle\langle1| +(-1)^m |2\rangle\langle2| +(-1)^m e^{in\pi}|3\rangle\langle3|, \nonumber  \\
U_{1n}^{m} = |1\rangle\langle0| + e^{in\pi}|0\rangle\langle1| +(-1)^m |3\rangle\langle2| +(-1)^m e^{in\pi}|2\rangle\langle3|, \nonumber  \\
U_{2n}^{m} = |2\rangle\langle0| + e^{in\pi}|3\rangle\langle1| +(-1)^m |0\rangle\langle2| +(-1)^m e^{in\pi}|1\rangle\langle3|, \nonumber  \\
U_{3n}^{m} = |3\rangle\langle0| + e^{in\pi}|2\rangle\langle1| +(-1)^m |1\rangle\langle2| +(-1)^m e^{in\pi}|0\rangle\langle3|. \nonumber  \\
\end{eqnarray}
Subsequently, photon is returned to the receiver, who can decode the classical information encoded by the sender with HDBSM for four-dimensional path entangled photons. In theory, if the complete four-dimensional BSM is realizable, the channel capacity of superdense coding can up to $log_216=4$ bits. However, as we know, the 16 states described in Equ. (1) cannot be unambiguously distinguished from each other via just linear optics. In the following text, we will demonstrate two theoretical approaches for linear-optical four-dimensional BSM in path DOF, resorting to the two-photon interference. 

\begin{figure}
\centering
\includegraphics*[width=0.35\textwidth]{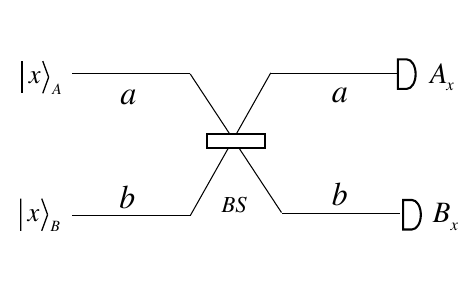}
\caption{Schematic diagram of the setup for linear-optical HDBSM using two-photon interference. The beam splitter (BS) performs the Hadamard operation [$|x\rangle_a\rightarrow(|x\rangle_a + |x\rangle_b)/{\sqrt 2},|x\rangle_b\rightarrow(|x\rangle_a - |x\rangle_b)/{\sqrt 2}; x = 0, 1, 2, 3$] on photonic path state. Photon-number resolving detectors are required to detect the more than one responses. With this device, the 16 four-dimensional Bell states in path DOF can be classified into 7 distinguishable groups.}
\end{figure}

The schematic diagram of our first approach is shown in Fig. 1. Before deriving the evolution results of the 16 four-dimensional Bell states, we first calculate the evolution of the following three different types of quantum states, which can be conveniently utilized for obtaining the final detection results. The evolution of state $|x\rangle_A|x\rangle_B (x=0,1,2,3)$ after passing BSs is
\begin{eqnarray}
|x\rangle_A|x\rangle_B \rightarrow |x\rangle_a|x\rangle_a - |x\rangle_b|x\rangle_b.
\end{eqnarray}
It is easy to find that the two identical photons will occur in the same output, and the photon-number resolving detectors (PNRDs) are required to recognize the more than one responses in one output. The evolution of state $|x\rangle_A|y\rangle_B + |y\rangle_A|x\rangle_B (0\leq x < y \leq3)$ after passing BSs is
\begin{eqnarray}
|x\rangle_A|y\rangle_B + |y\rangle_A|x\rangle_B \rightarrow |x\rangle_a|y\rangle_a - |x\rangle_b|y\rangle_b.
\end{eqnarray}
We can find that the photons will be detected in different outputs with the same $a$ or $b$, but not same $|x\rangle$ and $|y\rangle$. The evolution of state $|x\rangle_A|y\rangle_B - |y\rangle_A|x\rangle_B ((0\leq x < y \leq3)$ after passing BSs is
\begin{eqnarray}
|x\rangle_A|y\rangle_B - |y\rangle_A|x\rangle_B \rightarrow |y\rangle_a|x\rangle_b - |x\rangle_a|y\rangle_b.
\end{eqnarray}
Also, the two photons will be detected in different outputs. However, for this time, they are in the different $a$ and $b$, and different $|x\rangle$ and $|y\rangle$.

Overall, these three different types of quantum states can be completely discriminated from each other with the setup presented in Fig. 1. With this conclusion, the evolution of state $|\psi_{0n}^{m}\rangle$ can be written as
\begin{eqnarray}
&&|\psi_{0n}^{m}\rangle \rightarrow (|0\rangle_a|0\rangle_a - |0\rangle_b|0\rangle_b) + e^{in\pi}(|1\rangle_a|1\rangle_a - |1\rangle_b|1\rangle_b) \nonumber  \\ &&+ (-1)^m [(|2\rangle_a|2\rangle_a - |2\rangle_b|2\rangle_b) + e^{in\pi}(|3\rangle_a|3\rangle_a - |3\rangle_b|3\rangle_b)].  \nonumber  \\
\end{eqnarray}
These four states with the same parity information but the different relative phase information cannot be distinguished with this approach, and the detection results correspond to the Group 1 in Table 1. The evolution of states $|\psi_{10}^{m}\rangle$ and $|\psi_{11}^{m}\rangle$ can be written as 
\begin{eqnarray}
|\psi_{10}^{m}\rangle \rightarrow (|0\rangle_a|1\rangle_a - |0\rangle_b|1\rangle_b) + (-1)^m (|2\rangle_a|3\rangle_a - |2\rangle_b|3\rangle_b),  \nonumber  \\
|\psi_{11}^{m}\rangle \rightarrow (|1\rangle_a|0\rangle_b - |0\rangle_a|1\rangle_b) + (-1)^m (|3\rangle_a|2\rangle_b - |2\rangle_a|3\rangle_b).  \nonumber  \\
\end{eqnarray}
Therefore, states $|\psi_{10}^{m}\rangle$ and $|\psi_{11}^{m}\rangle$ are distinguishable with this approach, corresponding to the Group 2 and Group 3 in Table 1. Likely, the other 8 four-dimensional Bell states $|\psi_{2n}^{m}\rangle$ and $|\psi_{3n}^{m}\rangle$ can be classified into four groups with the similar evolution.

\begin{table}
\centering\caption{Detection result table. The 16 orthogonal four-dimensional path Bell states are measured by using two-photon interference.}
\begin{tabular}{cc ccccccccccc ccccccccccc}
\hline
Group & States & Detection results  \\
\hline
$1$ & 
$|\psi_{00}^{0}\rangle$, $|\psi_{00}^{1}\rangle$,& $A_0A_0, A_1A_1, A_2A_2, A_3A_3,$ \\ &
$|\psi_{01}^{0}\rangle$, $|\psi_{01}^{1}\rangle$ & $B_0B_0, B_1B_1, B_2B_2, B_3B_3.$ \\
$2$ & 
$|\psi_{10}^{0}\rangle$, $|\psi_{10}^{1}\rangle$ & $A_0A_1, B_0B_1, A_2A_3, B_2B_3.$ \\ 
$3$ & 
$|\psi_{11}^{0}\rangle$, $|\psi_{11}^{1}\rangle$ & $A_0B_1, A_1B_0, A_2B_3, A_3B_2.$ \\ 
$4$ & 
$|\psi_{20}^{0}\rangle$, $|\psi_{21}^{0}\rangle$ & $A_0A_2, B_0B_2, A_1A_3, B_1B_3.$ \\ 
$5$ & 
$|\psi_{20}^{1}\rangle$, $|\psi_{21}^{1}\rangle$ & $A_0B_2, A_2B_0, A_1B_3, A_3B_1.$ \\ 
$6$ & 
$|\psi_{30}^{0}\rangle$, $|\psi_{31}^{1}\rangle$ & $A_0A_3, B_0B_3, A_1A_2, B_1B_2.$ \\ 
$7$ & 
$|\psi_{30}^{1}\rangle$, $|\psi_{31}^{0}\rangle$ & $A_0B_3, A_3B_0, A_1B_2, A_2B_1.$ \\

\hline
\end{tabular}
\end{table}

As shown in Table 1, the 16 initial states can be divided into 7 distinguishable groups, based on the detection results of PNRDs. Thus, with this approach for HDBSM, the four-dimensional path Bell state can be utilized for superdense coding to transmit $log_27=2.81$ bits classical information with sending one photon. From Table 1, it is easy to find that PNRDs are only required to detect the high-dimensional states in the first group. Therefore, the channel capacity of superdense coding will decrease to $log_26=2.59$ bits, if we just exploit the common single-photon detector.

By introducing the additional two-dimensional polarization entanglement, we find that the more four-dimensional path Bell states can be determined. A quantum hyperentangled system that consists of four-dimensional path entanglement and two-dimensional polarization entanglement can be expressed as
\begin{eqnarray}
|\Psi\rangle_{AB} = \frac{1}{2}(|00\rangle + |11\rangle + |22\rangle + |33\rangle) \nonumber  \\ \otimes \frac{1}{\sqrt 2}(|HH\rangle + |VV\rangle)_{AB}.
\end{eqnarray} 
Here $|H\rangle$ and $|V\rangle$ are the horizontal and vertical polarization states of photon, respectively. It should be noted that this type of $2\times4$-dimensional hyperentanglement is achievable with the current quantum technology \cite{10,23,24}. With the auxiliary polarization state $|\phi\rangle=\frac{1}{\sqrt 2}(|HH\rangle + |VV\rangle)_{AB}$, the efficiency of our second approach for linear-optical HDBSM can be promoted, the schematic diagram of which is shown in Fig. 2.

\begin{figure}
\centering
\includegraphics*[width=0.45\textwidth]{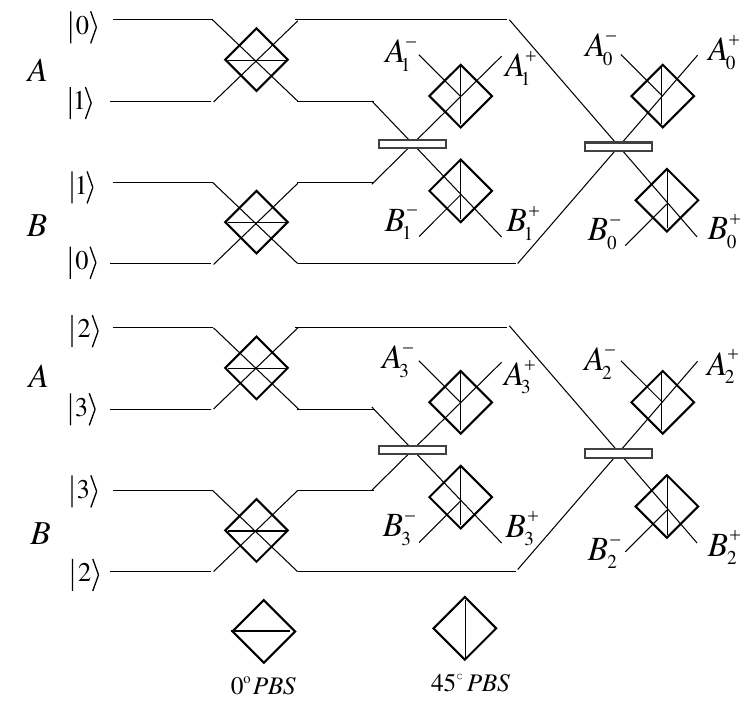}
\caption{Schematic diagram of the setup for linear-optical HDBSM using two-photon interference and auxiliary two-dimensional polarization entanglement. The polarization beam splitter (PBS) orientated at $0^{\circ}$ transmits the horizontal polarization photon and reflects the vertical one, while the PBS orientated at $45^{\circ}$ transmits state $|+\rangle=(|H\rangle + |V\rangle)/{\sqrt 2}$ and reflects state $|-\rangle=(|H\rangle - |V\rangle)/{\sqrt 2}$. With this device, the 16 four-dimensional path Bell states can be classified into 12 distinguishable groups.}
\end{figure}

Here, we use state $|\psi_{21}^{0}\rangle\otimes|\phi\rangle$ as the example to demonstrate the principle of Fig. 2, and the evolution of this state is as follows,
\begin{eqnarray}
&&|\psi_{21}^{0}\rangle\otimes|\phi\rangle \nonumber  \\&&= \frac{1}{2\sqrt 2}(|02\rangle - |13\rangle + |20\rangle - |31\rangle) \otimes (|HH\rangle + |VV\rangle)_{AB} \nonumber  \\ && \xrightarrow[0^\circ]{PBS} \frac{1}{2\sqrt 2}(|02\rangle - |13\rangle + |20\rangle - |31\rangle) \otimes (|HH\rangle - |VV\rangle)_{AB} \nonumber \\ &&\xrightarrow{BS}\frac{1}{2\sqrt 2}(|0\rangle_a|2\rangle_a - |0\rangle_b|2\rangle_b - |1\rangle_a|3\rangle_a + |1\rangle_b|3\rangle_b)\nonumber\\ &&\otimes (|HH\rangle - |VV\rangle)_{AB} \nonumber  \\ &&\xrightarrow[45^\circ]{PBS} \frac{1}{2\sqrt 2} (|A_0^+A_2^-\rangle + |A_0^-A_2^+\rangle - |A_1^+A_3^-\rangle - |A_1^-A_3^+\rangle \nonumber  \\&& - |B_0^+B_2^-\rangle - |B_0^-B_2^+\rangle + |B_1^+B_3^-\rangle + |B_1^-B_3^+\rangle).
\end{eqnarray}
This is the sixth group in Table 2, which shows the relations between the initial states and detection results. According to the detector outcomes, the 16 four-dimensional path Bell states can be classified into 12 groups with PNRDs, which can support the $log_212=3.58$ bits channel capacity for superdense coding. From Table 2, we can find that only the detection of states in the first group requires the PNRDs. Therefore, if PNRDs are not utilized, the channel capacity will decrease to $log_211=3.45$ bits. 

\begin{table}
\centering\caption{Detection result table. The 16 orthogonal four-dimensional path Bell states are measured by using two-photon interference and auxiliary two-dimensional polarization entanglement.}
\begin{tabular}{cc ccccccccccc ccccccccccc}
\hline
Group & States & Detection results  \\
\hline
$1$ & 
$|\psi_{00}^{0}\rangle$, $|\psi_{00}^{1}\rangle$ & $A_0^+A_0^+, A_1^+A_1^+, A_2^+A_2^+, A_3^+A_3^+,$ \\ && $B_0^+B_0^+, B_1^+B_1^+, B_2^+B_2^+, B_3^+B_3^+,$ \\ && $A_0^-A_0^-, A_1^-A_1^-, A_2^-A_2^-, A_3^-A_3^-,$ \\&& $B_0^-B_0^-, B_1^-B_1^-, B_2^-B_2^-, B_3^-B_3^-.$ \\
$2$ & 
$|\psi_{01}^{0}\rangle$, $|\psi_{01}^{1}\rangle$ & $A_0^+A_0^-, A_1^+A_1^-, A_2^+A_2^-, A_3^+A_3^-,$ \\ && $B_0^+B_0^-, B_1^+B_1^-, B_2^+B_2^-, B_3^+B_3^-.$ \\
$3$ & 
$|\psi_{10}^{0}\rangle$, $|\psi_{10}^{1}\rangle$ & $A_0^+A_1^+, A_0^-A_1^-, A_2^+A_3^+, A_2^-A_3^-,$ \\ && $B_0^+B_1^+, B_0^-B_1^-, B_2^+B_3^+, B_2^-B_3^-.$ \\
$4$ & 
$|\psi_{11}^{0}\rangle$, $|\psi_{11}^{1}\rangle$ & $A_0^+B_1^-, A_0^-B_1^+, A_1^+B_0^-, A_1^-B_0^+,$ \\ && $A_2^+B_3^-, A_2^-B_3^+, A_3^+B_2^-, A_3^-B_2^+.$ \\
$5$ & 
$|\psi_{20}^{0}\rangle$ & $A_0^+A_2^+, A_0^-A_2^-, A_1^+A_3^+, A_1^-A_3^-,$ \\ && $B_0^+B_2^+, B_0^-B_2^-, B_1^+B_3^+, B_1^-B_3^-.$ \\
$6$ & 
$|\psi_{21}^{0}\rangle$ & $A_0^+A_2^-, A_0^-A_2^+, A_1^+A_3^-, A_1^-A_3^+,$ \\ && $B_0^+B_2^-, B_0^-B_2^+, B_1^+B_3^-, B_1^-B_3^+.$ \\
$7$ & 
$|\psi_{20}^{1}\rangle$ & $A_0^+B_2^+, A_2^+B_0^+, A_1^+B_3^+, A_3^+B_1^+,$ \\ && $A_0^-B_2^-, A_2^-B_0^-, A_1^-B_3^-, A_3^-B_1^-.$ \\
$8$ & 
$|\psi_{21}^{1}\rangle$ & $A_0^+B_2^-, A_2^+B_0^-, A_1^+B_3^-, A_3^+B_1^-,$ \\ && $A_0^-B_2^+, A_2^-B_0^+, A_1^-B_3^+, A_3^-B_1^+.$ \\
$9$ & 
$|\psi_{30}^{0}\rangle$ & $A_0^+A_3^+, A_0^-A_3^-, A_1^+A_2^+, A_1^-A_2^-,$ \\ && $B_0^+B_3^+, B_0^-B_3^-, B_1^+B_2^+, B_1^-B_2^-.$ \\
$10$ & 
$|\psi_{31}^{1}\rangle$ & $A_0^+A_3^-, A_0^-A_3^+, A_1^+A_2^-, A_1^-A_2^+,$ \\ && $B_0^+B_3^-, B_0^-B_3^+, B_1^+B_2^-, B_1^-B_2^+.$ \\
$11$ & 
$|\psi_{30}^{1}\rangle$ & $A_0^+B_3^+, A_3^+B_0^+, A_1^+B_2^+, A_2^+B_1^+,$ \\ && $A_0^-B_3^-, A_3^-B_0^-, A_1^-B_2^-, A_2^-B_1^-.$ \\
$12$ & 
$|\psi_{31}^{0}\rangle$ & $A_0^+B_3^-, A_3^+B_0^-, A_1^+B_2^-, A_2^+B_1^-,$ \\ && $A_0^-B_3^+, A_3^-B_0^+, A_1^-B_2^+, A_2^-B_1^+.$ \\

\hline
\end{tabular}
\end{table}

\section{Discussion and summary}
The traditional BSM schemes that utilize two-photon interference is using the two-dimensional quantum entanglement of photon system, such as two-dimensional path Bell state and two-dimensional polarization Bell state \cite{17,18}. However, limited by the maximal discrimination efficiency of linear-optical BSM, the four orthogonal two-dimensional Bell states only can be classified into three groups, which is able to support the $log_23=1.58$ bits channel capacity in quantum dense coding \cite{17,18}. When an auxiliary DOF is introduced, the complete two-dimensional BSM can be accomplished \cite{19,20,21,22}, and thus a theoretical $log_24=2$ bits channel capacity becomes possible \cite{3}. Based on the principle of two-photon interference with two-dimensional Bell state, in this paper, we have theoretically investigated the accessible avenues for HDBSM using two-photon interference with four-dimensional Bell state. Owing to the larger information capacity of four-dimensional entanglement when compared with two-dimensional entanglement, our results are distinctly superior to the previous quantum dense coding protocols with quantum interference \cite{17,18,21,22}. Specifically speaking, on the condition of using no additional quantum resource, the channel capacity of 2.81 bits that supported by our first HDBSM approach is larger than the previous 1.58 bits \cite{17,18}. When an additional DOF is exploited, the channel capacity of 3.58 bits that supported by our second HDBSM approach is larger than the theoretical 2 bits \cite{3,21,22}. Actually, assisted by additional polarization DOF, the four-dimensional Bell state has already been experimentally utilized for superdense coding in 2018 and 2019, with the channel capacities of 2.09 bits and 3 bits, respectively \cite{9,10}. Compared with these two important results, the 3.58 bits channel capacity we achieved is higher, which will be significant for the future practical superdense coding based on high-dimensional hyperentanglement.

In summary, we have proposed two efficient approaches for the linear-optical HDBSM of photonic four-dimensional Bell state in path DOF using two-photon interference, and have demonstrated their applications in superdense coding. With only linear optics, our first approach can achieve a 2.81 bits/photon classical information transmission rate, and our second approach can promote the information transmission rate up to 3.58 bits/photon with an auxiliary two-dimensional polarization Bell state. Moreover, both of the principle and procedure of our linear-optical HDBSM approaches are clear, simple and realizable. We hope these two HDBSM approaches can be demonstrated in the experimental quantum platform, and can be helpful for other quantum communication protocols, quantum computation, quantum sensing and quantum imaging.



\end{document}